\documentclass[12pt]{article}
\usepackage{amssymb}

\def\be{\begin{equation}}
\def\ee{\end{equation}}
\def\bea{\begin{eqnarray}}
\def\eea{\end{eqnarray}}
\def\ba{\begin{array}}
\def\ea{\end{array}}
\def\bi{\begin{itemize}}
\def\ei{\end{itemize}}
\def\half{{\textstyle{1\over2}}}

\catcode`\@=11
\def\@citex[#1]#2{%
\if@filesw \immediate \write \@auxout {\string \citation {#2}}\fi
\@tempcntb\m@ne \let\@h@ld\relax \def\@citea{}%
\@cite{%
  \@for \@citeb:=#2\do {%
    \@ifundefined {b@\@citeb}%
      {\@h@ld\@citea\@tempcntb\m@ne{\bf ?}%
      \@warning {Citation `\@citeb ' on page \thepage \space undefined}}%
      {\@tempcnta\@tempcntb \advance\@tempcnta\@ne%
      \@tempcntb\number\csname b@\@citeb \endcsname \relax%
      \ifnum\@tempcnta=\@tempcntb 
        \ifx\@h@ld\relax%
          \edef \@h@ld{\@citea\csname b@\@citeb\endcsname}%
        \else%
          \edef\@h@ld{\ifmmode{-}\else--\fi\csname b@\@citeb\endcsname}%
        \fi%
      \else
        \@h@ld\@citea\csname b@\@citeb \endcsname%
        \let\@h@ld\relax%
      \fi}%
    \def\@citea{,\penalty\@highpenalty\,}%
  }\@h@ld
}{#1}}

\def\@citeb#1#2{{[#1]\if@tempswa , #2\fi}}
\def\@citeu#1#2{{$^{#1}$\if@tempswa , #2\fi }}
\def\@citep#1#2{{#1\if@tempswa , #2\fi}}

\def\bcites{         
        \catcode`\@=11
        \let\@cite=\@citeb
        \catcode`\@=12
}

\def\upcites{         
        \catcode`\@=11
        \let\@cite=\@citeu
        \catcode`\@=12
}

\def\plaincites{      
        \catcode`\@=11
        \let\@cite=\@citep
        \catcode`\@=12
}

\newcount\hour
\newcount\minute
\newtoks\amorpm
\hour=\time\divide\hour by 60
\minute=\time{\multiply\hour by 60 \global\advance\minute by-\hour}
\edef\standardtime{{\ifnum\hour<12 \global\amorpm={am}%
        \else\global\amorpm={pm}\advance\hour by-12 \fi
        \ifnum\hour=0 \hour=12 \fi
        \number\hour:\ifnum\minute<10 0\fi\number\minute\the\amorpm}}
\edef\militarytime{\number\hour:\ifnum\minute<10 0\fi\number\minute}

\def\draftlabel#1{{\@bsphack\if@filesw {\let\thepage\relax
   \xdef\@gtempa{\write\@auxout{\string
      \newlabel{#1}{{\@currentlabel}{\thepage}}}}}\@gtempa
   \if@nobreak \ifvmode\nobreak\fi\fi\fi\@esphack}
        \gdef\@eqnlabel{#1}}
\def\@eqnlabel{}
\def\@vacuum{}
\def\marginnote#1{}
\def\draftmarginnote#1{\marginpar{\raggedright\scriptsize\tt#1}}
\def\draft{
        \pagestyle{plain}
        \overfullrule=2pt
        \oddsidemargin -.5truein
        \def\@oddhead{\sl \phantom{\today\quad\militarytime} \hfil
        \smash{\Large\sl DRAFT} \hfil \today\quad\militarytime}
        \let\@evenhead\@oddhead
        \let\label=\draftlabel
        \let\marginnote=\draftmarginnote
        \def\ps@empty{\let\@mkboth\@gobbletwo
        \def\@oddfoot{\hfil \smash{\Large\sl DRAFT} \hfil}
        \let\@evenfoot\@oddhead}
        \def\@eqnnum{(\theequation)\rlap{\kern\marginparsep\tt\@eqnlabel}%
        \global\let\@eqnlabel\@vacuum}  }

\begin{document}


\hfill UTHET-02-0401

\vspace{-0.2cm}

\begin{center}
\Large
{ \bf Holography in the Penrose limit of AdS space\footnote{Research supported by the DoE under grant DE-FG05-91ER40627.}}
\normalsize

\vspace{0.8cm}
 
{\bf George Siopsis}\footnote{gsiopsis@utk.edu}\\ Department of Physics
and Astronomy, \\
The University of Tennessee, Knoxville, \\
TN 37996 - 1200, USA.
 \end{center}

\vspace{0.8cm}
\large
\centerline{\bf Abstract}
\normalsize
\vspace{.5cm}

We discuss the Penrose limit of pure AdS space, which is flat Minkowski space.
Even though there is no holographic principle, we construct a ``holographic
screen'' on which information on the corresponding CFT is encoded.
The screen is obtained as a gauge-fixing condition upon restricting the Hilbert
space to the states that are annihilated by the generator of scale
transformations. This constraint leads to Dirac brackets which turn the
Poincar\'e algebra into the algebra of the conformal group on the
``holographic screen.''

\newpage

The plane-wave limit of spacetime suggested by Penrose~\cite{bib1} a while ago and its supersymmetric generalization~\cite{bibsu} has recently
attracted a lot of attention in conjunction with the AdS/CFT correspondence~\cite{bibst1,bib4a,bib4aa,bib4ab,bib4,bib4e,bib4f,bib4g,bib4h,bib4i,bib4j,bib5,bib5a,bib7,bib7a,bib2,bib3,bib3a,bib3b,bib6}.
This limit leads to a more direct understanding of the AdS/CFT correspondence
because the background is simple enough for the sigma model (string theory)
to be exactly solvable~\cite{bibst1,bib4a,bib4aa,bib4ab,bib4,bib4e,bib4f,bib4g,bib4h,bib4i,bib4j,bib5,bib5a}. Other exactly solvable models studied earlier can be
shown to be special cases of the Penrose limit~\cite{bibpr0,bibpr1,bibpr2,bibpr3,bibpr4}.

On the other hand, the issue of holography is blurred in this limit, because
the background is a plane wave or even flat Minkowski space on which no
holographic principle exists. This is in contrast to the well-understood
holography in the original AdS space\footnote{For a review, see~\cite{bib9}.}
whose limit we are taking. Since the AdS/CFT correspondence seems to have survived
the limit, one is tempted to conjecture the existence of a holographic screen
on which the CFT resides. A number of proposals have been put forth~\cite{bib8,bib8aa,bib8a}, but the
issue remains open~\cite{bib8b}.

In this short note, we consider the simplest case of AdS$_{d+1}$ space whose Penrose
limit is flat Minkowski space $\mathcal{M}_{d+1}$. This may be viewed as a special case of the
Penrose limit of AdS$_{d+1}\times$S$^q$ in which we boost along a geodesic with
vanishing spin in the sphere S$^q$. One may define string theory on this
Minkowski space and study the CFT correspondence in a standard fashion.
Since no holographic principle applies to Minkowski space, it is evident
that an additional condition is needed for some kind of holography to emerge.
We introduce such a condition by restricting the Hilbert space to the
states that are scale invariant. This is a constraint leading to the treatment
of scale transformations as gauge transformations. In the resulting theory,
one needs to fix the gauge. This can be done by restricting to a
$d$-dimensional hypersurface
of the Minkowski space $\mathcal{M}_{d+1}$. This hypersurface is arbitrary as long as it cuts all
gauge orbits exactly once and all such choices are gauge equivalent.
We consider explicitly a hyperboloid (dS$_d$ space\footnote{Supersymmetry is
broken on dS$_d$, but it gets restored in the flat space limit.}) and show that the Poincar\'e
algebra of Minkowski space turns into a $SO(d,2)$ algebra once the Poisson
brackets are replaced by Dirac brackets. In the flat-space limit of the
hyperboloid, the $SO(d,2)$ group turns into the conformal group of the
resulting $d$-dimensional Minkowski space $\mathcal{M}_d$. This is then the
``holographic screen'' on which information on the corresponding CFT resides.

We start by reviewing the salient features of AdS space and the CFT
correspondence.
A $(d+1)$-dimensional AdS space (AdS$_{d+1}$)
is defined within a flat $(d+2)$-dimensional space as the hypersurface
\be\label{eqc1}
X_0^2 - X_1^2 - \dots - X_d^2 + X_{d+1}^2 = R^2
\ee
There are different choices of coordinate systems on the AdS hypersurface,
all of which should lead to the same physical results. However, one needs
to be careful, especially when taking the Penrose limit. The most
convenient choice are the Poincar\'e coordinates which, however, only cover
half of AdS.
The AdS$_{d+1}$ metric in Poincar\'e coordinates is
\be\label{eqmetric} ds^2 = R^2\, \frac{dr^2}{r^2} + \frac{r^2}{R^2} (-dt^2 + {d\vec x}^2)\ee
where $\vec x\in \mathbb{R}^{d-1}$.
It is also useful to introduce the invariant distance between
points $X^\mathbf{A}$ and ${X'}^\mathbf{A}$ ($\mathbf{A}=0,1,\dots,d+1$) on the AdS hypersurface,
\be\label{eqP} P(X,X') = \frac{1}{R^2}\; (X'-X)^2 =
\frac{(r-r')^2}{rr'}+\frac{rr'}{R^4}\; \Big( -(t-t')^2 +(\vec x-{\vec x}')^2\Big)\ee
In what follows, we shall use bold capitals as indices that take on $d+2$
values ($0,1,\dots,d+1$), plain capitals will run $M,N,\dots = 0,1,\dots,d$
and Greek indices will span a $d$-dimensional space ($\mu,\nu,\dots =
0,1,\dots,d-1$).
The group of isometries is $SO(d,2)$ whose generators we shall denote by
$J_\mathbf{AB}$.
The quadratic Casimir is
\be\label{eqqC} C_2 = \half J^\mathbf{AB} J_\mathbf{AB}\ee
For definiteness, we shall concentrate on scalar fields. The treatment
of fields of higher spin is similar. The scalar wave equation is
\be\label{eqwe}
\frac{1}{r^{d-1}}\; \frac\partial{\partial r} \left( r^{d+1} \frac{\partial\Psi}{\partial r}\right) - \partial_t^2\, \Psi +\vec\nabla^2\, \Psi = m^2R^2\, \Psi
\ee
The inner product in the space of solutions is given by
\be\label{eqip} (\Psi_1,\Psi_2) = i\pi R^4\;\int_\Sigma d^{d-1}x\,dr\,
r^{d-3} (\Psi_1^\star \partial_t
\Psi_2 - \partial_t \Psi_1^\star \Psi_2) \ee
where $\Sigma$ is the spacelike slice $t=$~const..
Assuming the wavefunction is a plane wave in the space spanned by $x^\mu =
(t, \vec x)$,
\be\label{eqpw} \Psi_k (r,x^\mu) = e^{ik_\mu x^\mu} \Phi_q (r)\ee
where $q^2= -k_\mu k^\mu$, we obtain
the solution to the wave equation in terms of Bessel functions
\be\label{eqphi}
\Phi_q^\pm (r) = r^{-d/2}\; J_{\pm\nu} (qR^2/r)\quad,\quad \nu = \half \sqrt{d^2 +4m^2R^2}\ee
The inner product of two wavefunctions is
\be\label{eqnorm} (\Psi_k^\pm,\Psi_{k'}^\pm) = (2\pi)^d \delta^{d-1}(\vec k - \vec k')
\; \frac{k_0+k_0'}{2}\; R^4\;
\int_0^\infty dr\, r^{d-3}\; \Phi_q^{\pm\star} (r)\Phi_{q'}^\pm (r)\ee
where $\vec k = (k_1,\dots, k_{d-1})$ and similarly for $\vec k'$.
At the boundary ($r\to \infty$), the two solutions behave as
\be\label{eqdim} \Phi_q^\pm \sim r^{-h_\pm} \quad,\quad h_\pm = \frac{d}{2} \pm \nu\ee
For $m^2>0$, the solution $\Phi_q^-$ is not normalizable, so it is discarded.
The normalizable modes form an orthonormal set with respect to the inner product~(\ref{eqip}),
\be (\Psi_k^+,\Psi_{k'}^+) = (2\pi)^d\, \delta^d(k -k')
\ee
where we used the orthogonality property of Bessel functions,
\be \int_0^\infty \frac{dr}{r^3}\, J_\nu (qR^2/r) \, J_\nu (q'R^2/r) = \frac{1}{qR^4}\, \delta (q-q')\ee
Next, we introduce the propagator
\be\label{eqgpsi} G (r,x^\mu ;r',{x'}^\mu) = \int \frac{d^d k}{(2\pi)^d}\, \Psi_k^{+\star} (r,x^\mu)\Psi_k^+ (r',{x'}^\mu)\ee
which obeys the wave equation~(\ref{eqwe}).
\footnote{We replaced the integration variable $q$ with $k_0$ in order to
arrive at a more convenient expression for the measure in~(\ref{eqgpsi}).}
After some algebra involving Bessel and Hypergeometric function identities,
we arrive at
\be\label{eqG}
G(z,x^\mu; z', {x'}^\mu) = \frac{\Gamma (h_+)}
{2\pi^{d/2}\Gamma (\nu+1)} \; P^{-h_+}\, F(h_+, \nu+\half; 2\nu+1;-4/P)\ee
where the invariant distance $P$ is given by~(\ref{eqP}).
The singularity is obtained by letting $P\to 0$,
\be\label{eqsing} G(z,x^\mu; z', {x'}^\mu) \sim \frac{\Gamma((d-1)/2)}
{4\pi^{(d+1)/2}} \; P_\epsilon^{-(d-1)/2}\ee
where $P_\epsilon$ includes the $i\epsilon$ prescription $x^0-{x'}^0 \to
x^0-{x'}^0 -i\epsilon$.
The bulk-to-boundary propagator is obtained by letting one of the arguments
approach the boundary.
In the limit $r'\to 0$, we have $G(r,x^\mu; r', {x'}^\mu) \sim P^{-h_+}$, so
\be G(r,x^\mu; r' {x'}^\mu) \to \frac{1}{2\nu}\; {r'}^{-h_+}\; K(r,x^\mu;{x'}^\mu)\ee
where
\be\label{eqgf} K(r,x^\mu;{x'}^\mu) = \pi^{-d/2}\; \frac{\Gamma(h_+)}{\Gamma (\nu)}\;
r^{h_+}\; \left( 1+\frac{r^2}{R^4}\; (x-x')^\mu (x-x')_\mu\right)^{-h_+}\ee
In the limit $r\to \infty$, this leads to a propagator of the form
\be\label{eqgf1} \Delta (x) \sim (x^\mu x_\mu)^{-h_+}\ee
which is the two-point function (up to a constant) of the corresponding conformal field theory.

Next, we wish to study the Penrose limit of the theory. To this end, we shall
boost along a null geodesic. All choices are equivalent the most convenient being
the radial direction.
A radial null geodesic is given by ($ds^2=0$)
\be\label{eqg1} R^2\; \frac{\dot r^2}{r^2} - \frac{r^2}{R^2}\; \dot t^2 =0\ee
where we differentiate with respect to the affine parameter $\tau$. Conservation
of energy (independence of the metric on $t$)
implies
\be\label{eqt} \frac{r^2}{R^2}\; \dot t = E\ee
where $E$ is a constant of the motion. By rescaling $\tau$, we may set $E=1$.
Then the equation for the geodesic~(\ref{eqg1}) becomes
\be\label{eqg2} \dot r^2 = 1\ee
so $r = -\tau$ may be chosen as the affine parameter.
Along the geodesic, we may choose
\be t = \frac{R^2}{r}\ee
which is the solution of eq.~(\ref{eqt}).
Next we shift the coordinate $t$ by its value on the geodesic,
\be t = \frac{R^2}{r} -\tilde t\ee  
In terms of the new coordinate $\tilde t$, the AdS metric~(\ref{eqmetric}) reads
\be ds^2 = -2drd\tilde t-\frac{r^2}{R^2}\; d\tilde t^2 +\frac{r^2}{R^2}\; {d\vec x}^2\ee
In the Penrose limit, this becomes
\be ds_P^2 = -2drd\tilde t +\frac{r^2}{R^2}\; {d\vec x}^2\ee
This is flat Minkowski space. To see this, change variables to
\be \vec y = \frac{r}{R}\; \vec x\quad,\quad v = \tilde t +\frac{r}{2R^2}\; {\vec x}^2\ee
The metric becomes
\be ds_P^2 = -2drdv + {d\vec y}^2\ee
which may be brought into the standard form
\be ds_P^2 = dY^M dY_M = -(dY^0)^2 + dY^idY^i\ee
where $Y^i = y^i\;\; (i=1,\dots, d-1)$, $Y^0 = \frac{1}{\sqrt 2} (r+v)$ and $Y^d = \frac{1}{\sqrt 2} (r-v)$.
%
It is easily seen that the invariant distance~(\ref{eqP}) turns into the
Minkowski space distance in the Penrose limit,
\be P(X,X') \to \frac{1}{R^2}\; (Y-Y')^M(Y-Y')_M\ee
Since $R\to \infty$ in this limit, the only contribution to the Green function
(\ref{eqG}) that survives is the small distance singularity~(\ref{eqsing}),
\be G(P) \sim P^{-(d-1)/2}\ee
Consequently, the boundary scaling behavior~(\ref{eqgf}), which is obtained
from the large $P$ limit, is not present in the Minkowski space~\cite{bibgidd}. The group
of isometries $SO(d,2)$ is mapped onto the Poincar\'e group,
\be M_{MN} = J_{MN}\quad,\quad P_M = \frac{J_{M(d+1)}}{R}\quad\quad
(M,N = 0,1,\dots,d)\ee
Indeed, it is readily deduced from the $SO(d,2)$ algebra,
\bea
[M_{MN}\;,\; M_{PQ}] &=& -i\eta_{MP} M_{NQ} + \dots \nonumber \\
\; [ M_{MN}\;,\; P_P] &=& -i (\eta_{MP} P_N -\eta_{NP} P_M)\nonumber \\
\; [ P_M\;,\; P_N] &=& 0\label{eqPa}
\eea
as $R\to\infty$. The quadratic Casimir~(\ref{eqqC})
becomes
\be C_2 \to R^2 P_MP^M = -m^2R^2\ee
Therefore, its value does not change. Since the parameter $m^2R^2$ is still
present in the Penrose limit, one might wonder whether the scaling dimensions
$h_\pm$ (\ref{eqdim}), which are solely functions of $m^2R^2$, have survived.
At first glance, it appears that no conclusion on scaling dimensions may be
drawn from the Casimir $C_2$, for its value comes from the abelian
part of the Poincar\'e algebra~(\ref{eqPa}).
On the other hand, one may define and solve a sigma model (string theory)
on the flat Minkowski space and recover the scaling dimensions from CFT
correlators. Thus, even though holography is absent on Minkowski space,
information on the CFT on the boundary of AdS space has not been lost.
It is natural to look for a ``holographic screen'' on which this information
may reside.

To this end, let us concentrate on the operators which are scale invariant.
Correspondingly, we restrict attention to the subspace of states which are
annihilated by the generator of scale transformations,
\be \Delta |\Psi\rangle = 0\quad,\quad \Delta = Y^M P_M\ee
Once this is imposed as a second-class constraint, scale transformations become
gauge transformations. Then we must fix the gauge in the resulting theory.
This is accomplished by restricting our space to a $d$-dimensional hypersurface. A convenient choice (gauge-fixing condition) is
\be Y^MY_M = R^2\ee
which is $d$-dimensional
de Sitter space (dS$_{d}$). The Poisson brackets then turn into
Dirac brackets,
\be \{ \mathcal{A}\;,\; \mathcal{B}\}_D = \{ \mathcal{A}\;,\; \mathcal{B}\}_P
-\frac{1}{2R^2}\; \Big( \{\mathcal{A}\;,\;\chi_1\}_P\{\chi_2\;,\;\mathcal{B}\}_P
- \{\mathcal{A}\;,\;\chi_2\}_P\{\chi_1\;,\;\mathcal{B}\}_P\Big)\ee
where $\chi_1 = \Delta$, $\chi_2 = Y^MY_M$, and we used
$\{ \chi_1\;,\; \chi_2\}_P = 2R^2$.

The Poincar\'e algebra~(\ref{eqPa}) is modified as follows. The Lorentz
group generators $M_{MN}$ commute with the constraints, so their algebra
does not change. By the same token, nor do the commutators of a Lorentz
group generator and a momentum $P_M$. The Dirac bracket of two momentum
components is
\be \{ P_M\;,\; P_N\}_D = \frac{M_{MN}}{R^2}\ee
Therefore, their abelian algebra is modified,
\be [P_M\;,\; P_N] = -i\; \frac{M_{MN}}{R^2}\ee
This shows that $RP_M$ can be identified with the $SO(d,2)$ generator
$J_{M(d+1)}$ and the modified algebra is $SO(d,2)$. In the flat space limit
($R\to\infty$, around, e.g., the $Y^d$ axis), the dS$_{d}$ space becomes $d$-dimensional Minkowski space. To see what becomes of the $SO(d,2)$ algebra in
this limit, let us parametrize the dS hypersurface using global coordinates
$(\tau, \vec y)$, where $\vec y = (y_1,\dots, y_{d-1})$, as
\be Y^0 = R\sinh\tau\quad,\quad Y^i = R\cosh\tau\; y_i\quad,\quad
Y^d = R\cosh\tau\; \sqrt{1-{\vec y\, }^2}\ee
The dS metric reads
\be\label{eqdsmet} ds_{dS}^2 = -R^2 d\tau^2 + R^2\cosh^2\tau \left( {d\vec y\, }^2 +
\frac{(\vec y\cdot d\vec y)^2}{1-{\vec y\, }^2} \right)\ee
The $SO(d,2)$ generators may also be expressed in terms of the dS coordinates.
A short calculation yields
$$ RP_0 = \cosh\tau\; p_\tau\quad,\quad
RP_i = -\sinh\tau\; y_i p_\tau
+\frac{(\delta_{ij} - y_iy_j)}{\cosh\tau} p_j
$$
\be RP_d = -\sqrt{1-{\vec y\, }^2}\left(\sinh\tau\; p_\tau
-\frac{y_j}{\cosh\tau} p_j\right)\ee
$$M_{0i} = -y_i \left( p_\tau - \tanh\tau y_j p_j \right) - \tanh\tau p_i
\quad,\quad M_{ij} = y_i p_j -y_j p_i$$
\be M_{0d} = -\sqrt{1-{\vec y\, }^2}\left(p_\tau
+\tanh\tau y_j\, p_j\right)
\quad,\quad M_{id} = -\sqrt{1-{\vec y\, }^2} \; p_i\ee
where $i,j=1,\dots, d-1$ and we have defined the conjugate momenta
\be\label{eqppy} p_\tau = i \frac{\partial}{\partial\tau}\quad,\quad p_i = -i \frac{\partial}{\partial y_i}\ee
By scaling
\be \tau \to \tau /R\quad,\quad y_i\to y_i/R\ee
in the limit $R\to\infty$, the metric~(\ref{eqdsmet}) becomes
\be ds_{R\to\infty}^2 = dy_\mu dy^\mu = - (dy^0)^2 + {d\vec y\, }^2\ee
where $y^0 = \tau$,
showing explicitly that the dS hypersurface has turned into flat $d$-dimensional Minkowski space
spanned by coordinates $y^\mu = (y^0,\dots, y^{d-1})$.
Ignoring higher-order terms in each of the $SO(d,2)$ generators,
as $R\to\infty$, we obtain
$$ P_0 = \left(1+\frac{\tau^2}{2R^2} \right)\; p_\tau\quad,\quad
P_i = -\frac{\tau y_i}{R^2} p_\tau
+\left(1-\frac{\tau^2}{2R^2}\right) p_i
- \frac{y_iy_j}{R^2} p_j
$$
\be RP_d = y^\mu p_\mu\ee
$$M_{\mu\nu} = y_\mu p_\nu -y_\nu p_\mu$$
\be \frac{M_{0d}}{R} = -\left(1-\frac{{\vec y\, }^2}{2R^2}\right) p_\tau
+\frac{\tau y_j}{R^2}\, p_j
\quad,\quad \frac{M_{id}}{R} = -\left(1-\frac{{\vec y\, }^2}{2R^2}\right) \; p_i\ee
The $SO(d,2)$ group is then the conformal group on the $d$-dimensional
Minkowski space in the $R\to\infty$ limit through
the identifications
\be \mathcal{M}_{\mu\nu} = M_{\mu\nu}\quad,\quad
\mathcal{K}_\mu = 2R^2\; \left( P_\mu + \frac{M_{\mu d}}{R}\right)\quad,\quad 
\mathcal{P}_\mu = \half \left( P_\mu - \frac{M_{\mu d}}{R}\right)\quad,\quad
\mathcal{D} = RP_d
\ee
Explicitly,
\be\mathcal{M}_{\mu\nu} = y_\mu p_\nu -y_\nu p_\mu\quad,\quad \mathcal{P}_\mu = p_\mu
\quad,\quad \mathcal{D} = y^\mu p_\mu\quad,\quad
\mathcal{K}_\mu = y^2p_\mu -2y_\mu y^\nu p_\nu\ee
where $p_\mu = i\partial / \partial y^\mu$ (eq.~(\ref{eqppy})).
A scalar operator of conformal dimension $h$ satisfies
\bea
[\mathcal{M}_{\mu\nu}\;,\; \mathcal{A} (y)] &=& i(y_\mu\partial_\nu
-y_\nu\partial_\mu) \mathcal{A} (y)\nonumber\\
\, [\mathcal{P}_\mu\;,\; \mathcal{A} (y)] &=& i\partial_\mu\mathcal{A} (y)\nonumber\\
\, [\mathcal{D}\;,\; \mathcal{A} (y)] &=& i(y^\mu\partial_\mu -h)\mathcal{A} (y)\nonumber\\
\, [\mathcal{K}_\mu\;,\; \mathcal{A} (y)] &=& i(y^2\partial_\mu -2y_\mu y^\nu\partial_\nu +2y_\mu h)\mathcal{A} (y)\eea
where $\partial_\mu = \partial / \partial y^\mu$.
The quadratic Casimir~(\ref{eqqC}) is
\be C_2 = \half \mathcal{M}_{\mu\nu} \mathcal{M}^{\mu\nu} + \half (\mathcal{K}_\mu \mathcal{P}^\mu
+ \mathcal{P}_\mu \mathcal{K}^\mu ) - \mathcal{D}^2 = -h(d-h) \ee
Setting it equal to $-m^2R^2$, we obtain the conformal weights
\be h = h_\pm = \half (d\pm\sqrt{d^2+4m^2R^2})\ee
in agreement with the AdS result~(\ref{eqdim}). Unitarity selects $h_+$ as the only
possible weight. The conformal group also determines
all correlators of the primary field $\mathcal{A}$ up to constants. In
particular, the two-point function is
\be \langle \mathcal{A} (y) \mathcal{A} (y')\rangle \sim ((y-y')^2)^{-h_+}\ee
in agreement with the holographic result~(\ref{eqgf1}) from AdS space.
Thus, we have recovered a conformal field theory on the $d$-dimensional Minkowski space
which was obtained as the flat-space limit of a dS$_d$ hypersurface in the
original $(d+1)$-dimensional Minkowski space. The restriction on dS$_d$
induced the replacement of Poisson brackets by Dirac brackets.
Consequently, the $(d+1)$-dimensional
Poincar\'e algebra turned into a $SO(d,2)$ algebra which was identified
with the conformal group on the $d$-dimensional Minkowski space. It should be emphasized that this
``holographic screen'' is merely a gauge choice and therefore arbitrary.

It would be interesting to include spin and extend the above results to the
Penrose limit of AdS$\times$S which is a pp-wave. This will be reported elsewhere.

\section*{Acknowledgements}

I wish to thank J.~T.~Liu and A.~C.~Petkou for discussions.

\end{document}